\documentclass{PoS}

\newcommand{\bra}{\left \langle}
\newcommand{\ket}{\right \rangle}
\newcommand{\eq}{\begin{equation}} 
\newcommand{\en}{\end{equation}} 

\def\one{{\rm 1\kern -.9mm l}}                             %

\newcommand{\eqa}{\begin{eqnarray}}
\newcommand{\ena}{\end{eqnarray}}
\def\beq{\begin{equation}}
\def\eeq{\end{equation}}
\def\beq{\begin{equation}}
\def\eeq{\end{equation}}
\def\beqa{\begin{eqnarray}}
\def\eeqa{\end{eqnarray}}

\def\ee{\mathrm{e}}                                                           
\newcommand{\Z}{\mathbb{Z}}

\title{New results on the effective string corrections to the inter-quark potential.}

\ShortTitle{New results on the effective string corrections to the inter-quark potential.}

\author{{Marco Bill\`o},~~ \speaker{Michele Caselle},~~ Valentina Verduci~ and~ Mirco Zago\\
        Dipartimento di Fisica, Universit\`a di Torino \\
        and Istituto Nazionale di Fisica Nucleare, sezione di Torino\\
        via P. Giuria 1, 10125 Torino (Italy)\\
        E-mail: \email{(billo)(caselle)(verduci)(zago)@to.infn.it}}

\abstract{We propose a new approach to the study of the inter-quark potential in Lattice Gauge Theories.
Instead of looking at the expectation value of Polyakov loop correlators we study the modifications
induced in the chromoelectric flux by the presence of the Polyakov loops. In abelian LGTs, thanks to
duality, this study can be performed in a very efficient way, allowing to reach high precision with a 
reasonable CPU cost. The major advantage of this numerical strategy is that it allows to eliminate the 
dominant effective string correction to the inter-quark potential (the L\"uscher term) thus giving an 
unique opportunity to test higher order corrections. Performing a set of simulations in the $3d$ gauge 
Ising model we were thus able to precisely identify and measure both the quartic and the sextic 
effective string corrections to the inter-quark potential. While the quartic term perfectly agrees with 
the Nambu-Goto one the sextic term is definitely different. Our result seems to disagree with the recent 
proof by Aharony and Karzbrun of the universality of the sextic correction. We discuss a few possible 
explanations of this disagreement. 

The numerical approach described above can also be applied to the study of Wilson loops. In this case, 
the numerical results  are precise enough to test the two-loop prediction of the Nambu-Goto action. The 
two-loop NG result computed time ago by by Dietz and Filk is incompatible with the data; however, after 
correcting some mistakes in their expression, compatibility is restored. The viability of a 
first-order, operatorial description of the Wilson loop is also pointed out.}

\FullConference{The XXVIII International Symposium on Lattice Field Theory\\
	 June 14-19,2010\\
     Villasimius, Sardinia Italy}

\begin{document}

One of the most interesting recent results in the context of the effective string approach to LGT's is the proof of universality of  
the first few terms of the effective action.  Universality up to the quartic
order was proved a few years ago by L\"uscher and Weisz~\cite{lw04} and recently extended to the sixth 
order by Aharony and Karzbrun \cite{Aharony:2009gg} (see also \cite{newlist} for related results). These
findings are based on a set of assumptions and it would 
be of great importance to test them with a numerical simulation. This is a very difficult task for at least two reasons: 
\begin{itemize} 
\item
 in the standard "zero temperature" inter-quark potential, higher order corrections are proportional to higher powers in $1/R$ ($R$ being the inter-quark distance) and are thus visible only at very short distance, where the effective string picture breaks down and spurious effects (due for
 instance to boundary terms) and perturbative contributions become important;
\item
 the dominant string correction (the so called L\"uscher term)  may shadow the sub-leading terms.
 \end{itemize}

In order to overcome these two problems we propose the following strategy. First, instead of working at zero temperature we shall study the 
inter-quark potential at finite temperature (just below the deconfinement transition). It is well known that in this regime the string corrections  are proportional to $R$
and act as a temperature dependent renormalization of the string tension. 
In this regime higher order corrections correspond to higher powers of $T$ and can be observed much
better than in the zero temperature limit.
Second, in order to eliminate the dominant L\"uscher term we shall not measure directly
the inter-quark potential, but the changes induced in the flux configuration by the presence of the Polyakov loops. We shall show below that 
as a consequence of this choice the L\"uscher term does not contribute. This makes this approach a very
efficient tool to explore higher order corrections. 


The lattice operator which measures the flux through a plaquette $p$ in presence of two Polyakov loops $P$, $P'$ is:
\eq 
\bra\phi(p;P,P')\ket=\frac{\bra P P'^\dagger~U_p\ket}{\bra PP'^\dagger \ket}-\bra U_p\ket~,
\label{flux2} 
\en 
where $U_p$ is, as usual, the trace of the ordered product of link variables along the plaquette $p$.
We shall be interested in the following to the mean flux density, defined as:
\eq
\bra\Phi(R,L)\ket=\frac{1}{N_p}\sum_p \frac{\bra P P'^\dagger~U_p\ket}{\bra PP'^\dagger \ket}-\bra U_p\ket~,
\en
where $N_p$ denotes the number of plaquettes of the lattice, $R$ is the spatial distance between
the Polyakov loops, namely, the inter-quark distance, and $L$ the the number of lattice spacings
in the time-like direction. In the following we will also introduce  
$N_s$ to indicate the number of lattice spacings in the space-like 
directions, so that in $d=2+1$ we have $N_p=3N_s^2L$.

It is easy to see that if we define the partition function of the system in presence of the two Polyakov loops as 
\eq
Z(R,L)=\langle P^\dagger (R) P(0) \rangle 
\en
then $\bra\Phi(R,L)\ket$ can be written as:
\eq
\bra\Phi(R,L)\ket=\frac{1}{N_p}\frac{d}{d \beta} \log Z(R,L)~,
\label{defPhi}
\en
where $\beta$ is the coupling appearing in the Wilson action with respect to which the expectation values are being taken. If we neglect for the moment effective string corrections and keep only the area term in $Z$, i.e. $Z(L,R)\sim \ee^{-\sigma RL}$
we find a linearly rising behaviour for $\bra\Phi(R,L)\ket$:
\eq
\bra\Phi(R,L)\ket=\alpha R
\label{Phi2}
\en
with an angular coefficient which in (2+1) dimensions is given by
\eq
\alpha=\frac{L}{N_p}\frac{d \sigma}{d \beta} =\frac{1}{3N_s^2}\frac{d \sigma}{d \beta}
\label{defalpha}
\en
and does not depend on the finite temperature $1/L$.  

The effective string corrections to $Z(L,R)$ 
depend on the particular string action that we choose. They can be expanded   
in powers of the dimensionless quantity $(\sigma RL)^{-1}$:

\eq
Z(L,R)=\ee^{-\sigma RL} \cdot Z_{1} \cdot \left( 1+ \frac{F_4}{\sigma RL} + \frac{F_6}{(\sigma RL)^2} +\cdots \right)~,
\label{zexp}
\en
where the indices in $F_4$ and $F_6$ recall the fact that they are obtained from the quartic and sextic terms in the expansion of the effective string action respectively.

The leading order of this expansion, namely $Z_1$,  corresponds to the partition function of a free boson in two dimensions. This term is universal and does not depend on the string tension $\sigma$. 
Its dominant contribution in the large $L$ limit is the well known 
"L\"uscher term"~\cite{Luscher:1980fr}.  The fact that $Z_1$ is scale invariant implies that it must also be $\beta$ independent and thus disappears in eq. (\ref{defPhi}). As anticipated above, 
the effective string corrections to $\bra\Phi(R,L)\ket$ start at the first sub-leading term, the "quartic"  correction $F_4$.
In the Nambu-Goto case all the terms $F_n$ are known \cite{lw04,Billo:2005iv}.
Inserting eq. (\ref{zexp}) in eq. (\ref{defPhi}) we obtain the effective string corrections to $\bra\Phi(R,L)\ket$. Since we shall be
interested in the large $R$ limit of these corrections, we find convenient to organize them via an expansion in powers of $1/R$: 
\eq
\bra\Phi(R,L)\ket~=~
\alpha 
\left(R A(x) + B(x) + \frac{C(x)}{R} + ...\right)~, 
\en
where $x\equiv\frac{\pi}{3\sigma L^2}$. In the Nambu-Goto case these functions can be evaluated explicitly \cite{cz2010} and turn out to be:
\eq
A(x)=\frac{(1-x/2)}{\sqrt{1-x}}
\label{ng1}
\en
\eq
B(x)=\frac{1}{4\sigma L}\frac{x}{(1-x)}~,
\en
\eq
C(x)=\frac{1-x/2}{8(L\sigma)^2(1-x)^{3/2}}~.
\label{ng3}
\en
While $\alpha$ depends on the details of the gauge theory  and in particular it reflects the specific $\beta$ dependence of the string tension $\sigma$, the functions $A(x), B(x), C(x), \ldots$ encode the information on the effective string model. 
The particular form of these functions given in eq.s (\ref{ng1}-\ref{ng3})
corresponds to the Nambu-Goto model; however, according to~\cite{lw04,Aharony:2009gg}, the first two orders in the perturbative expansion in power of $1/\sigma$ (i.e. in
powers of $x$) should be universal. In order to identify these terms, let us expand 
these functions in powers of $x$ :
\eq
 A(x)= \left(1+\frac{x^2}{8}+\frac{x^3}{8}+...\right)~,
\label{def1}
\en
\eq
 B(x)=\frac{1}{\sigma L}\left(\frac{x}{4}+\frac{x^2}{4}+...\right)~
\label{def2}
\en
\eq
 C(x)=\frac{1}{8(\sigma L)^2}\left(1+ x+ \ldots\right)~.
\label{def3}
\en


To test these corrections we performed a set of high precision simulations in the $3d$ gauge Ising model, using the same methods discussed in \cite{Allais:2008bk,cgmv95}. 
We mapped via duality the Polyakov loops correlator
into the partition function of a $3d$ Ising spin model in which we changed the sign of the coupling 
of all the links dual to the surface bordered by the two Polyakov loops.
We then estimated $\bra\Phi(R,L)\ket$ by simply evaluating the mean energy in presence of these frustrated links. Further  details on the algorithm and on the results of the
simulations can be found in~\cite{cz2010}.
Since duality plays a crucial role in this derivation, the approach discussed in this paper is particularly suited for abelian gauge theories; given enough computational power, however,
it could be extended to non-abelian models.

We chose to simulate the model at $\beta=0.75180$, for which both $\sigma$ and the deconfinement temperature are known with very high precision: $\sigma=0.0105241(15)$ from \cite{Caselle:2004jq}
and  $1/T_c=L_c=8$  from~\cite{Caselle:1995wn}.
Using the above values for $\sigma$ and $T_c$ and keeping into account the scaling correction as discussed in \cite{Caselle:2007yc}  we obtain for the prediction that  $\alpha=2.792~~10^{-5}$  (see \cite{cz2010} for a detailed derivation). 
To test at the same time this prediction and the form of the effective string correction
we selected two sets of values of $L$. 
The first set contains the  values $L=16,20,24$ and
the variable $x$  was tuned so that the sextic term $x^3/8$ in eq. (\ref{def1}) is negligible with respect to the errors, but the quartic one, $x^2/8$, is not. The second set instead comprises   $L=10,11,12$; in these cases also the sextic term is definitely not negligible with respect to the errors.

For each value of L we fitted the data for $\Phi(R,L)$ according to the law
\eq
\Phi(R,L)=a(L)R+b(L)+c(L)/R~,
\en 
where the term $c(L)/R$ was introduced only for the second set of values of $L$ because it was always compatible with zero in the first set. 
In the following we shall concentrate on the values of $a(L)$ which are the most precise and allow to perform a stringent test of the effective string prediction.

The values of $a(L)$ extracted from the fits are reported in tab.\ref{tab4} and plotted in fig.1. We analyzed these data in two steps. First 
we fitted the first three values of $a(L)$ (those corresponding to $L=16,20,24$) with the law:
\eq
a(L)=\alpha(1+\gamma x^2)~.
\en
We obtained $\alpha=2.7918(17)~10^{-5}$ and  $\gamma= 0.132(7)$ with a very good reduced $\chi^2$. Both these values nicely agree with the predictions $\alpha=2.792~~10^{-5}$
and $\gamma=1/8$. 

We then fitted the  whole set of data (i.e. including also $L=10,11,12$) with the law:
\eq
a(L)=\alpha(1+\gamma x^2+\delta x^3)~.
\en
We found again a good reduced $\chi^2$:  $\chi^2_r=0.45$ and the following best fit results:
$$ \alpha=2.796(5)~10^{-5},~~~ \gamma= 0.127(25),~~~ \delta= -0.051(27)~. $$
The first two values agree again very well with the predictions but the coefficient of the sextic correction, which should be $\delta=1/8$, completely disagrees with them. 
This can be well appreciated looking at fig.1, where we plotted the quartic correction (dashed curve), the sextic correction according to the Nambu-Goto effective action (dashed-dotted curve)
and the curve corresponding to the best fit value of the parameter $\delta$ (dotted curve).

\begin{table}[htpb]
\centering
\begin{tabular}{|c|c|c|c|c|}
  \hline
  $L$ & $a(L)$  &  &  &      \\
  \hline\hline
  10  & 3.017(21)  & 2.792 & 3.137   & 3.481      \\
  11  & 2.943(20)  & 2.792 & 3.027   & 3.222   \\
  12  & 2.917(13)  & 2.792 & 2.959   & 3.074  \\
  16  & 2.847(9)  & 2.792 & 2.845    & 2.865 \\
  20  & 2.816(9)  & 2.792 & 2.814    & 2.819 \\
  24  & 2.802(8)  & 2.792 & 2.802   & 2.804 \\
  \hline
\end{tabular}
\caption{\small \textit{Values of the coefficient $a(L)$ 
for various values of $L$. In the second column we list the results of the simulations extracted from the fits to eq.(15)
In the following columns we 
report the prediction for $a(L)\equiv \alpha A(x)$ at the zeroth, first and second order in the expansion in $x$}}
\label{tab4}
\end{table}

\begin{figure}[htpb]
  \hskip -1cm
  \includegraphics[width=5 in]{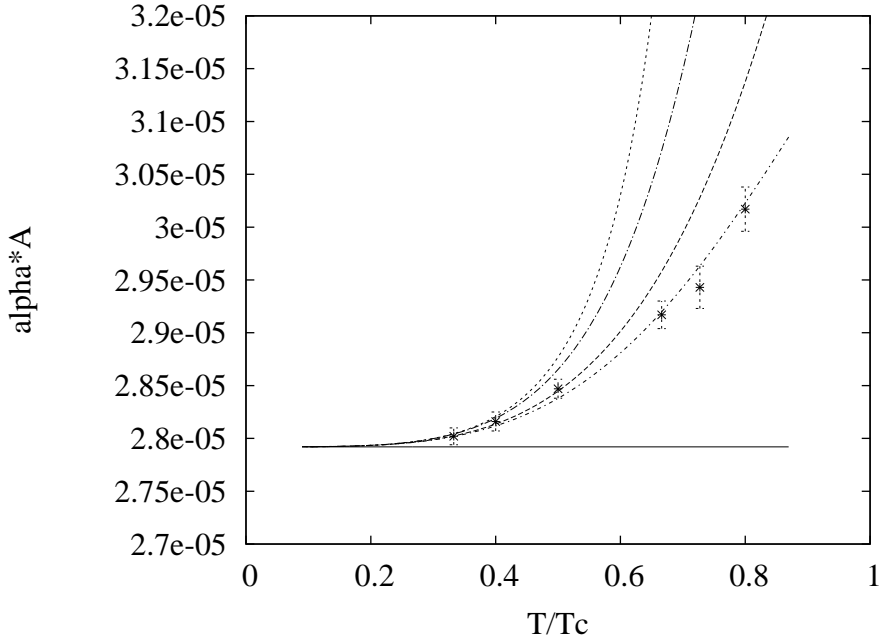}
  \caption{\small \textit{Plot of $\alpha~A(L)$ as a function of $\frac{T}{T_c}=\frac{8}{L}$. The continuous line is the prediction for 
  $\alpha$ (i.e. without effective string corrections). The other curves correspond, from top to bottom to the whole Nambu-Goto prediction, eq.(9), 
  the truncation at the
  sextic order, the truncation at the quartic order and finally the dashed dotted line corresponds to our best fit result ($\delta= -0.05$) 
  for the sextic coefficient (see the text).
  The points are the results of the simulations in the $3d$ gauge Ising model (see tab.1)
  }}
  \label{fig1}
\end{figure}

This observation agrees with a set of similar results obtained in these last years in the $3d$ gauge Ising model by considering various physical observables,
ranging from the inter-quark potential to the effective string width to the interface free energy of the dual spin Ising
model \cite{chp03,Caselle:2004jq,Caselle:2007yc,Caselle:2010zs}. All these tests 
supported, even if only at a qualitative level, a value for the sextic correction different from the Nambu-Goto one and fully compatible with the one that we find here.

Our result is rather puzzling in view of the recent proof of the universality of the string effective action up to this order in $d=3$~\cite{Aharony:2009gg}. 
A possible explanation could be that the observed deviation is only a lattice artifact due to unusually large scaling corrections and that the correct sextic contribution is 
recovered in the continuum limit. A careful analysis of the scaling behaviour of $\bra\Phi(R,L)\ket$ is reported in~\cite{cz2010} and seems to exclude this scenario:
$\delta$ shows a mild dependence on $\beta$ as the critical point is approached. It remains always negative and seems to slowly converge toward the value $\delta=0$.

Another possibility is that the deviation signals that $3d$ gauge Ising model does not admit a weakly coupled effective string description. This weak coupling  limit is a basic assumption  of \cite{lw04,Aharony:2009gg,Billo:2005iv} and amounts to ask that the partition function of the string describing a particular surface (say
the cylindric surface connecting the two Polyakov loops) can be written as a sum of single string states propagating along the surface. 
The fact that this argument could be relevant for the $3d$ gauge Ising model is also supported by the intuitive observation that a gauge theory based on the 
$\Z_2$ group is indeed the farthest  possible choice with respect to the large $N$ limit of the SU$(N)$ gauge theory which is known to behave as a weakly coupled string theory. However if this is the reason behind the disagreement at the sixth order, it is not clear why the effective description works instead so well up to the quartic order. 
In order to better understand this issue we plan to perform the same analysis discussed here for other LGTs so as to gain some insight on the dependence of the $\delta$ parameter on the gauge group.

Another interesting application of our method is the study of the effective string corrections for Wilson loops. A long standing problem in this context is the fact that the Dietz and Filk~\cite{df83} result for the quartic correction in the case of the Nambu-Goto action is manifestly incompatible with the Arvis spectrum~\cite{Arvis:1983fp} for the open string. 
In particular, in the limit of very 
asymmetric Wilson loops, the Dietz and Filk result is one order of magnitude larger than (and opposite in sign to) the one suggested by the open string spectrum. 
To address this issue we performed a set of high precision simulations 
choosing the same values of $\beta$ discussed above and found a value for the quartic term perfectly compatible with
the Arvis spectrum while the Dietz and Filk result turned out to be excluded by more than ten standard deviations~\cite{bcv2010}. Triggered by this result we went through the original Dietz and Filk calculation and identified the origin of the discrepancy. The correct result turns out to be compatible both with the Arvis spectrum and with the numerical estimates. 
For a rectangular Wilson loop of sizes $R$ and $T$ in a $D+2$ dimensional LGT 
(i.e. with $D$ transverse directions) it can be written in terms of Eisenstein functions as follows~\cite{bcv2010}:
\eq
\frac{F_4}{\sigma RT}=\frac{1}{\sigma RT}\left(\frac{\pi}{24}\right)^2 \left[2D \left(\frac{T}{R}\right)^2 E_4\left(i\frac{T}{R}\right) ~ - ~ 
\frac{D(D-4)}{2} E_2\left(i\frac{T}{R}\right)E_2\left(i\frac{R}{T}\right) \right]~.
\label{2lng}
\en

The NG string corrections to the Wilson loop can also be computed along the lines used in \cite{Billo:2005iv} and \cite{interfaces} in the case of Polyakov loops 
and interfaces, respectively. Replacing the Nambu-Goto second-order description
with a first-order Polyakov one, the Wilson loop v.e.v. can be given an operatorial description: it corresponds to the propagation of a Dirichlet string whose end-points are attached to the spatial boundaries of the Wilson loop between two states which represent its emission and re-absorption at times $0$ and $T$ (such states were already considered in \cite{Imamura:2005zm}). This way we can obtain a closed expression for the Wilson loop which can be expanded in powers of $1/(\sigma R T)$, which in the NG framework corresponds to the loop expansion; with respect to the NG treatment, it is rather straightforward to extract higher loop contributions.
Preliminary computations show that at two loop eq. (\ref{2lng}) is reproduced up to the addition of a constant term in the square bracket; the precision of our numerical results does not allow to check if this constant term is supported by data or not.  
Such a (rather mild) discrepancy did not arise in the cases of Polyakov loops and interfaces, and it would certainly be very interesting to understand better its origin. In \cite{bcv2010} we plan to develop the operatorial approach
in a much more detailed way.


\end{document}